\documentclass[12pt]{article}
\usepackage{amsfonts}

\begin{document}

\def\theequation{\thesection.\arabic{equation}}

\bibliographystyle{unsrt}

\title{\bf Formal and Precise Derivation of the Green Functions for a Simple
Potential}

\author{R.~de la Madrid\,\thanks{Departamento de F{\'\i}sica Te\'orica, 
At\'omica y Nuclear, Universidad de Valladolid, Valladolid, Spain.} 
\thanks{Center for Particle Physics, The University of Texas at Austin, 
Austin, Texas, USA.}
\thanks{E-mail: \texttt{rafa@physics.utexas.edu}}} 

\date{March 15, 2001}

\maketitle

\begin{abstract}
In formal scattering theory, Green functions are obtained as solutions of a 
distributional equation.  In this paper, we use the Sturm-Liouville theory to 
compute Green functions within a rigorous mathematical theory.  We shall show 
that both the Sturm-Liouville theory and the formal treatment yield the same 
Green functions.  We shall also show how the analyticity of the Green
functions as functions of the energy keeps track of the so-called 
``incoming'' and ``outgoing'' boundary conditions.
\end{abstract}

PACS number(s): 03.65.Nk, 23.60.+e*

\def\thesection{\arabic{section}}
\section{Introduction}
\def\thesection{\arabic{section}}
\setcounter{equation}{0}
\label{sec:introduction}

Green functions are essential tools in mathematical 
physics~\cite{DIMITRIEVA, ECONOMU}.  Most books on scattering 
theory~\cite{NEWTON, TAYLOR} only
provide a computational procedure to obtain Green functions
without any reference to a mathematical setting.  Here we provide 
that mathematical framework for the example of a square barrier 
potential.  In this example, the mathematical framework is given by 
the Sturm-Liouville theory~\cite{DUNFORD}. 

We consider a square barrier potential of height $V_0$,
\begin{equation}
              V(r)=\left\{ \begin{array}{ll}
                                0   &0<r<a  \\
                                V_0 &a<r<b  \\
                                0   &b<r<\infty \, . 
                  \end{array} 
                 \right.  
	\label{potential}
\end{equation}
Due to the spherical symmetry of the potential, spherical coordinates come
in handy.  The expression for the Hamiltonian in spherical coordinates 
(for $s$-waves) is
\begin{equation}
       h \equiv - \frac{d^2}{dr^2}+V(r) \, .
      \label{doh}
\end{equation}
In order to obtain a linear operator from the formal differential 
operator~(\ref{doh}), we need to define a domain in the Hilbert space 
$L^2([0,\infty ),dr)$ on which $h$ acts.  The domain we choose is
\begin{equation}
       {\cal D}(H) =\{ f(r)\, | \ 
                   (hf)(r), 
                   f(r)\in L^2([0,\infty ), dr), \,
                   f(0)=0, \, f(r) \in AC^2[0,\infty ) \} \, .
     \label{domain}
\end{equation}
On this domain, the formal differential operator~(\ref{doh}) induces a
self-adjoint operator $H$ (cf.~\cite{DUNFORD}).  The spectrum of this 
operator is the positive real line $[0,\infty )$ (cf.~\cite{DUNFORD}).

Both formal and Sturm-Liouville approaches use the solutions of the  
time-independent Schr\"odinger equation,
\begin{equation}
      \left( -\frac{d^2}{dr^2}+V(r) \right) \sigma (r,E)=
      E\sigma (r,E) \, ,
      \label{rSe0}
\end{equation}
as their basic source of information.  The difference is that 
physicists, fearful of using complex energies when working with self-adjoint
Hamiltonians, solve (\ref{rSe0}) using real energies.  The Sturm-Liouville 
approach uses complex energies.  This is in no contradiction with the 
self-adjointness of the Hamiltonian, since the eigenfunctions that correspond
to complex energies lie outside the domain~(\ref{domain}) on which the 
Hamiltonian is self-adjoint.

In the next section, we show that the formal and the Sturm-Liouville 
approaches yield the same Green functions.  In our calculations, we will 
use the following branch of the square root function:
\begin{equation}
      \sqrt{\cdot}:\{ E\in \mathbb C \, | \  -\pi <{\rm arg}(E)\leq \pi \} 
   \longmapsto \{E\in \mathbb C \, | \  -\pi/2 <{\rm arg}(E)\leq \pi/2 \} \, . 
\end{equation}

\def\thesection{\arabic{section}}
\section{Computation of the Green Functions}
\def\thesection{\arabic{section}}
\setcounter{equation}{0}
\label{sec:CoGF}

\subsection{Formal Approach}
\label{sec:green}

To compute the (radial) Green functions, physicists solve the following 
distributional equation:
\begin{equation}
     \left( -\frac{d^2}{d r^2}+V(r)-E\right)
       G(r,s;E)=-\delta (r-s) \, , \quad E\in [0,\infty ) \, ,
      \label{greene}
\end{equation}
subject to certain boundary conditions.  Equation~(\ref{greene}) says that 
for $r\neq s$, $G(r,s;E)$ obeys the radial Schr\"odinger 
equation~(\ref{rSe0}) and the following boundary conditions:
\begin{eqnarray}
       G(a-,s;E)&=&G(a+,s;E) \label{bc1} \\
       G(b-,s;E)&=&G(b+,s;E) \\
       \frac{\partial}{\partial r}G(a-,s;E)&=&
        \frac{\partial}{\partial r}G(a+,s;E)  \\
       \frac{\partial}{\partial r}G(b-,s;E)&=&
        \frac{\partial}{\partial r}G(b+,s;E) \, . 
\end{eqnarray}
At $r=s$ it is continuous, but its derivative 
has a discontinuity of 1,
\begin{equation}
	\frac{\partial}{\partial r}G(s+0,s;E)-
	\frac{\partial}{\partial r}G(s-0,s;E)=1 \, ,
\end{equation}
due to the delta function. At the origin, the Green function must be 
regular,
\begin{equation}
       G(0,s;E)=0 \, . \label{bc8}
\end{equation}
      
There are two linearly independent Green functions that we are mostly 
interested in: the {\it incoming} and the 
{\it outgoing} Green functions.  The outgoing Green function $G^+$ satisfies 
Eq.~(\ref{greene}), the boundary conditions 
(\ref{bc1})-(\ref{bc8}), and an ``outgoing boundary condition'' at 
infinity,
\begin{equation}
      G^+(r,s;E)\sim e^{i\sqrt{E}r} \, , \quad {\rm for \ } r\to \infty \, .
       \label{obc}
\end{equation}
The boundary condition~(\ref{obc}) means that very far from the potential 
region the function $G^+$ behaves as an outgoing wave.  The expression for 
this outgoing Green function is~\cite{NEWTON}
\begin{equation}
	G^+(r,s;E)=\frac{\chi (r_<;E) \Omega _+(r_>;E)}{W(\chi ,\Omega _+ )}
        \, , \label{pogf}
\end{equation}
where $r_<$ and $r_>$ refer to the smaller and to the bigger of $r$ and $s$, 
respectively, $\chi (r;E)$ is the solution of the Schr\"odinger equation that 
vanishes at the origin, $\Omega _+(r;E)$ is the solution of the 
Schr\"odinger equation that satisfies an ``outgoing boundary condition'' 
at infinity, and $W(\chi ,\Omega _+ )$ is the the Wronskian of $\chi$ and 
$\Omega _+$.  More precisely, $\chi (r;E)$ satisfies Eq.~(\ref{rSe0}) and the 
boundary conditions
\begin{eqnarray}
        \chi (0;E)&=&0  \label{chi0} \\
	\chi (a+;E)&=&\chi (a-;E)  \label{chi1}  \\
	\chi '(a+;E)&=&\chi '(a-;E) \\
	\chi (b+;E)&=&\chi (b-;E)    \\
	\chi '(b+;E)&=&\chi '(b-;\;E) \, .   \label{chi4}
\end{eqnarray}
Its expression is given by
\begin{equation}
      \chi (r;E)=\left\{ \begin{array}{lll}
               \sin (\sqrt{E}r) \quad &0<r<a  \\
               {\cal J}_1(E)e^{i \sqrt{(E-V_0)}r}
                +{\cal J}_2(E)e^{-i\sqrt{(E-V_0)}r}
                 \quad  &a<r<b \\
               {\cal J}_3(E) e^{i\sqrt{E}r}
                +{\cal J}_4(E)e^{-i\sqrt{E}r}
                 \quad  &b<r<\infty \, .
               \end{array} 
                 \right. 
             \label{chi}
\end{equation}
The functions ${\cal J}_1$-${\cal J}_4$ are such that $\chi (r;E)$ satisfies
the boundary conditions 
(\ref{chi0})-(\ref{chi4}) and are given in the Appendix.  The 
function $\Omega _+(r;E)$ satisfies Eq.~(\ref{rSe0})
and the boundary conditions
\begin{eqnarray}
	\Omega _+ (a+;E)&=&\Omega _+ (a-;E)  \label{bc+1}  \\
	\Omega _+ '(a+;E)&=&\Omega _+ '(a-;E) \\
	\Omega _+(b+:E)&=&\Omega _+ (b-;E)    \\
	\Omega _+'(b+;E)&=&\Omega _+'(b-:\;E) \label{bc+4} \\
        \Omega _+(r;E) &\sim&e^{i\sqrt{E}r} \, , \ r\to \infty \,. \label{bc+5}
\end{eqnarray}
The solution $\Omega _+(r;E)$ reads (see the Appendix for
the explicit expressions of ${\cal A}^+_1$-${\cal A}^+_4$)
\begin{equation}
      \Omega _+(r;E)=\left\{ \begin{array}{lll}
               {\cal A}^+_1(E)e^{i\sqrt{E}r}
               +{\cal A}^+_2(E) e^{-i\sqrt{E}r} 
                \quad &0<r<a  \\
               {\cal A}^+_3(E)e^{i\sqrt{(E-V_0)}r}
                +{\cal A}^+_4(E)e^{-i\sqrt{(E-V_0)}r}
                 \quad  &a<r<b \\
               e^{i\sqrt{E}r}
                 \quad  &b<r<\infty \, . 
               \end{array} 
                 \right. 
               \label{O+}
\end{equation}
The Wronskian of $\chi (r;E)$ and $\Omega _+(r;E)$ is
\begin{equation}
       W(\chi , \Omega _+)=2i\sqrt{E} {\cal J}_4(E) \, .
\end{equation}
The outgoing Green function is therefore given by
\begin{equation}
	G^+(r,s;E)=\frac{\chi (r_<;E) \Omega _+(r_>;E)}
                       {2i\sqrt{E} {\cal J}_4(E)} \, .
\end{equation}

The incoming Green function $G^-$ satisfies Eq.~(\ref{greene}), the boundary 
conditions (\ref{bc1})-(\ref{bc8}), and an ``incoming boundary condition'' at 
infinity,
\begin{equation}
      G^-(r,s;E)\sim e^{-i\sqrt{E}r} \, , \quad {\rm for \ } r\to \infty \, .
      \label{ibc}
\end{equation}
Condition (\ref{ibc}) says that far away from the potential region $G^-$ 
behaves as an incoming wave.  The expression for this incoming Green 
function is
\begin{equation}
      G^-(r,s;E)=\frac{\chi (r_<;E) \Omega _-(r_>;E)}{W(\chi ,\Omega _- )}\,,
\end{equation}
where $\chi (r;E)$ is the solution of the Schr\"odinger equation that 
vanishes at the origin, $\Omega _-(r;E)$ is the solution of the 
Schr\"odinger equation that satisfies an ``incoming boundary condition'' 
at infinity, and $W(\chi ,\Omega _- )$ is the the Wronskian of $\chi$ and 
$\Omega _-$.  The eigenfunction $\chi (r;E)$ is given by (\ref{chi}).  The 
eigenfunction $\Omega _-(r;E)$ satisfies the Schr\"odinger equation subject to
the following boundary conditions:
\begin{eqnarray}
	\Omega _- (a+;E)&=&\Omega _- (a-;E) \label{O-1} \\
	\Omega _- '(a+;E)&=&\Omega _- '(a-;E) \\
	\Omega _-(b+:E)&=&\Omega _- (b-;E)    \\
	\Omega _-'(b+;E)&=&\Omega _-'(b-:\;E)  \\
        \Omega _-(r;E) &\sim&e^{-i\sqrt{E}r} \, , \ r\to \infty \,, \label{O-5}
\end{eqnarray}
and its expression is given by
\begin{equation}
      \Omega _-(r;E)=\left\{ \begin{array}{lll}
               {\cal A}_1^-(E)e^{i\sqrt{E}r}
               +{\cal A}_2^-(E) e^{-i\sqrt{E}r} 
                \quad &0<r<a  \\
               {\cal A}_3^-(E)e^{i\sqrt{(E-V_0)}r}
                +{\cal A}_4^-(E)e^{-i\sqrt{(E-V_0)}r}
                 \quad  &a<r<b \\
               e^{-i\sqrt{E}r}
                 \quad  &b<r<\infty \, . 
               \end{array} 
                 \right.
        \label{O-} 
\end{equation}
The functions ${\cal A}_1^-$-${\cal A}_4^-$ make $\Omega _-(r;E)$ satisfy
the boundary conditions (\ref{O-1})-(\ref{O-5}).  The Wronskian of 
$\chi (r;E)$ and $\Omega _-(r;E)$ is
\begin{equation}
       W(\chi , \Omega _-)=-2i\sqrt{E} {\cal J}_3(E) \, .
\end{equation}
Therefore, the expression of the incoming Green function is 
\begin{equation}
	G^-(r,s;E)=-\frac{\chi (r_<;E) \Omega _-(r_>;E)}
                       {2i\sqrt{E} {\cal J}_3(E)} \, .
        \label{pigf}
\end{equation}

\subsection{Sturm-Liouville Approach}
\label{sec:RaGF}

In this section, we compute the Green functions using 
the Sturm-Liouville theory.  The Green function appears as a kernel of 
integration when we write the resolvent of the Hamiltonian $H$ as an integral 
operator,
\begin{equation}
      \left( E-H \right) ^{-1}f(r)=
      \int_0^{\infty}G(r,s;E)f(s)\, ds \, , \quad E \notin [0,\infty ) \, .
       \label{MGF}
\end{equation}
From Eq.~(\ref{MGF}) one can formally derive Eq.~(\ref{greene}).

As mentioned in the Introduction, physicists avoid computing Green 
functions using complex energies, but rather impose the 
incoming and outgoing boundary conditions.  The Sturm-Liouville theory 
deals directly with complex energies in a way that keep track of the
``in'' and ``out'' boundary conditions.

The procedure to compute the Green function for our Hamiltonian $H$ 
is given by the following theorem (see Theorem XIII.3.16 in~\cite{DUNFORD}):

\vskip0.5cm

{\bf Theorem 1}  Let $H$ be the self-adjoint operator derived
from the real formal differential operator (\ref{doh}) by the imposition
of the boundary condition $f(0)=0$.  Let ${\rm Im}(E) \neq 0$.  Then there is 
exactly one solution $\chi (r;E)$ of $(h-E)\sigma =0$ square-integrable at
$0$ and satisfying the boundary condition $f(0)=0$, and exactly one 
solution $\Omega (r;E)$ of $(h-E)\sigma =0$ square-integrable at infinity.  The
resolvent $(E-H)^{-1}$ is an integral operator whose kernel $G(r,s;E)$ is
given by
\begin{equation}
       G(r,s;E)=\left\{ \begin{array}{ll}
                  \frac{\chi (r;E) \, \Omega (s;E)}{W(\chi ,\Omega )}
                     &r<s \\ [2ex]
                  \frac{\chi (s;E) \, \Omega (r;E)}{W(\chi ,\Omega )}
                       &r>s  \, .
                  \end{array} 
                 \right. 
	\label{exofGFA}
\end{equation}
The incoming and outgoing Green functions are defined by
\begin{equation}
       G^{\pm}(r,s;E)=\lim_{\mu \to 0}G(r,s;E\pm i\mu) \, .
       \label{boul}
\end{equation}

\vskip0.5cm

From Eq.~(\ref{boul}) it follows that the Green functions $G^{\pm}$ are just 
the boundary values on the real axis of the kernel (\ref{exofGFA}), which is
a function of complex variable.  Thus $G^{\pm}$ keep track of what 
is ``incoming'' and of what is ``outgoing'' by specifying which side of
the cut of the resolvent we are on.  In our example, the cut of the 
resolvent, i.e., the spectrum of the Hamiltonian, is the positive real 
axis.

First we compute the kernel (\ref{exofGFA}) in the region ${\rm Im} (E)>0$.  In
this region, the eigenfunction $\chi (r;E)$ in (\ref{exofGFA}) satisfies 
the boundary conditions
\begin{eqnarray}
       && \chi (0;E)=0 \, , \label{bca01} \\
       && \chi (r;E)\in AC^2([0,\infty )) \, , \label{bcACc}   \\
       && \chi (r;E)
           {\rm \ is \ square \ integrable \ at \ } 0 \, .
       \label{bca03}
\end{eqnarray}
The boundary condition (\ref{bcACc}) implies the boundary conditions
(\ref{chi1})-(\ref{chi4}), and (\ref{bca03}) is automatically 
fulfilled.  This means that the eigenfunction $\chi (r;E)$ of Theorem 1, which
is unique, is the same as the eigenfunction of Eq.~(\ref{chi}), although now 
$E$ is allowed to run over the upper half-plane of complex numbers.  

The eigenfunction $\Omega (r;E)$ of Theorem 1, that we denote by 
$\Omega _+(r;E)$ for the case of ${\rm Im} (E)>0$, satisfies 
the Schr\"odinger equation and the boundary conditions
\begin{eqnarray}
       &&\Omega _+(r;E)\in AC^2([0,\infty )) \, , \label{fivc+} \\
       &&\Omega _+ (r;E) \ 
         {\rm is \ square \ integrable \ at \ } \infty \, .
         \label{bcainfty3}
\end{eqnarray}
The boundary condition (\ref{fivc+}) implies the continuity 
conditions (\ref{bc+1})-(\ref{bc+4}).  Condition (\ref{bcainfty3}) is, for
${\rm Im}(E)>0$, equivalent to (\ref{bc+5}).  Thus the (unique) function 
$\Omega _+(r;E)$ of Theorem~1 coincides with (\ref{O+}), $E$ being now any
complex number with positive imaginary part.  Thus, for ${\rm Im} (E)>0$, the
expression of $G(r,s;E)$ in (\ref{exofGFA}) is given by
(\ref{pogf}), although now $E$ is allowed to be any complex number in the 
upper half-plane.

For ${\rm Im} (E)<0$, the situation is similar.  The eigenfunction $\chi (r;E)$
of Theorem~1 is also given by (\ref{chi}).  The other 
eigenfunction in (\ref{exofGFA}), that we denote by $\Omega _-(r;E)$, 
coincides with the eigenfunction of Eq.~(\ref{O-}).  Needless to say, $E$ 
can now be any complex number in the lower half-plane.  Therefore, the kernel 
(\ref{exofGFA}) is given by (\ref{pigf}), $E$ now being any complex number 
with negative imaginary part. 

If we now compute the limits in (\ref{boul}), we obtain exactly the same
outgoing and incoming Green functions as those obtained in the previous 
section. 

\def\thesection{\arabic{section}}
\section{Conclusions}
\def\thesection{\arabic{section}}
\setcounter{equation}{0}
\label{sec:Conclus}

We have seen that the Green functions obtained by applying the Sturm-Liouville
theory are the same as those used in formal scattering theory.  We have also
seen that in order to find out what is ``incoming'' and what is ``outgoing'', 
the Sturm-Liouville theory substitutes the standard physicists boundary 
conditions by statements on the analyticity of the Green functions with 
respect to the energy variable---knowing which side of the cut we are on is 
tantamount to knowing what is ``incoming'' and what is ``outgoing''.

These conclusions are not restricted to the square barrier 
potential.  Actually, they hold for potentials that decrease fast enough at 
the origin and at infinity, and that do not have too many discontinuities.

\def\thesection{\arabic{section}}
\section*{Acknowledgments}
\def\thesection{\arabic{section}}
\setcounter{equation}{0}
\label{sec:ACk}

The author wishes to express his gratitude to M.~Mithaiwala for proofreading
the paper and making many invaluable suggestions, and to Prof.~J.~P.~Antoine 
and Dr.~T.~Kuna for introducing him to the Sturm-Liouville theory.  Financial 
support from the Welch Foundation is gratefully acknowledged.

\def\thesection{\arabic{section}}
\section*{Appendix}
\def\thesection{\arabic{section}}
\setcounter{equation}{0}
\label{sec:A}

The functions that appear in the expressions of the eigenfunctions 
in Section~\ref{sec:green} are given by
\begin{eqnarray}
      &&{\cal J}_1(E)=\frac{1}{2}e^{-i\sqrt{(E-V_0)}a} 
     \left( \sin (\sqrt{E }a)+
      \frac{ \sqrt{E}}{i \sqrt{(E-V_0)}}
      \cos (\sqrt{E}a) \right) \qquad \\
      &&{\cal J} _2 (E)=\frac{1}{2}e^{i\sqrt{(E-V_0)}a} 
     \left( \sin (\sqrt{E }a)-
      \frac{ \sqrt{E }}{i \sqrt{(E-V_0)}}
      \cos (\sqrt{E}a) \right) \qquad \\
      &&{\cal J}_3(E)=\frac{1}{2}e^{-i\sqrt{E}b}
                   \left[
            \left(1+
      \frac{ \sqrt{(E-V_0)}}{\sqrt{E}}
       \right) e^{i \sqrt{(E-V_0)}b}{\cal J}_1(E) \right. \nonumber \\
      && \hskip4cm \left. +\left(1-
      \frac{ \sqrt{(E-V_0) }}{\sqrt{E}}
       \right) e^{-i \sqrt{(E-V_0)}b}{\cal J}_2(E) 
        \right] \qquad  \\
       &&{\cal J}_4(E)=\frac{1}{2}e^{i\sqrt{E}b}
                   \left[
            \left(1-
      \frac{ \sqrt{(E-V_0)}}{\sqrt{E}}
       \right) e^{i \sqrt{(E-V_0)}b}{\cal J} _1(E) \right. \nonumber \\
       && \hskip4cm \left. +\left(1+
      \frac{ \sqrt{(E-V_0) }}{\sqrt{E}}
       \right) e^{-i \sqrt{(E-V_0)}b}{\cal J}_2(E) 
        \right] \, , \qquad
\end{eqnarray}

\begin{eqnarray}
     &&{\cal A}_3^+(E)=\frac{1}{2}e^{-i\sqrt{(E-V_0)}b} 
     \left(1+\frac{\sqrt{E }}
      {\sqrt{(E-V_0)}}\right) 
     e^{i\sqrt{E}b} \qquad \\
     &&{\cal A}_4^+(E)=\frac{1}{2}e^{i\sqrt{(E-V_0)}b}
       \left(1-\frac{ \sqrt{E }}
            { \sqrt{(E-V_0)} } \right)
       e^{i\sqrt{E }b} \qquad \\
     &&{\cal A}_1^+(E)=\frac{1}{2}e^{-i\sqrt{E }a}
                   \left[
      \left(1+\frac{\sqrt{(E-V_0)}}
      {\sqrt{E}}\right) 
      e^{i\sqrt{(E-V_0)}a}{\cal A}_3^+(E)   \right. \nonumber \\
      && \hskip4cm  \left. +\left(1-\frac{ \sqrt{(E-V_0)}}
            { \sqrt{E} } \right) 
      e^{-i\sqrt{(E-V_0)}a} {\cal A}_4^+(E)
     \right] \qquad \\ 
     &&{\cal A}_2^+(E)=\frac{1}{2}e^{i\sqrt{E }a}
                   \left[
     \left(1-\frac{ \sqrt{(E-V_0)}}
            { \sqrt{E } } \right) 
     e^{i\sqrt{(E-V_0)}a}{\cal A}_3^+(E) \right. \nonumber \\
      && \hskip4cm \left. +\left(1+\frac{ \sqrt{(E-V_0)}}
            { \sqrt{E} } \right) 
      e^{-i\sqrt{(E-V_0)}b}{\cal A}_4^+(E)
     \right] \, . \qquad
\end{eqnarray}
\begin{eqnarray}
    &&{\cal A}_3^-(E)=\frac{1}{2}e^{-i\sqrt{(E-V_0)}b} 
     \left(1-\frac{\sqrt{E}}
      {\sqrt{(E-V_0)}}\right) 
     e^{-i\sqrt{E}b} \qquad \\
    &&{\cal A}_4^-(E)=\frac{1}{2}e^{i\sqrt{(E-V_0)}b}
       \left(1+\frac{ \sqrt{E}}
            { \sqrt{(E-V_0)} } \right)
       e^{-i\sqrt{E}b} \qquad \\
    &&{\cal A}_1^-(E)=\frac{1}{2}e^{-i\sqrt{E }a}
                   \left[
      \left(1+\frac{\sqrt{(E-V_0)}}
      {\sqrt{E}}\right) 
      e^{i\sqrt{(E-V_0)}a}{\cal A}_3^-(E)  \right. \nonumber \\
       && \hskip4cm \left. +\left(1-\frac{ \sqrt{(E-V_0)}}
            { \sqrt{E} } \right) 
      e^{-i\sqrt{(E-V_0)}a}{\cal A}_4^-(E)
     \right] \qquad \\ 
    &&{\cal A}_2^-(E)=\frac{1}{2}e^{i\sqrt{E}a}
                   \left[
     \left(1-\frac{ \sqrt{(E-V_0)}}
            { \sqrt{E} } \right) 
     e^{i\sqrt{(E-V_0) }a}{\cal A}_3^-(E)  \right. \nonumber \\
     && \hskip4cm \left. + \left(1+\frac{ \sqrt{(E-V_0)}}
            { \sqrt{E} } \right) 
      e^{-i\sqrt{(E-V_0)}a}{\cal A}_4^-(E)
     \right] \, . \qquad
\end{eqnarray}

\end{document}